\documentstyle[prl,aps,epsf,twocolumn]{revtex}
\input epsf
\begin{document}
\title{Lattice theory of trapping reactions with mobile species
}
\author{
\parindent=0.0in
{M.Moreau,$^1$ G.Oshanin,$^1$
O.B{\'e}nichou,$^2$ M.Coppey $^1$}
{\\
\textit{$^1$ Laboratoire de Physique Th{\'e}orique des Liquides,
University Pierre et Marie Curie, 75252 Paris Cedex 05, France \\
\vspace*{.1in}
$^2$ Laboratoire de Physique de la Mati\`ere Condens\'ee,
Coll{\`e}ge de France, 11 place Marcelin Berthelot, 75005, Paris, France}\\
}
(Submitted on April 24, 2003; accepted for publication in PRE on January 8, 2004)\\
\parbox{14cm}{
We present a stochastic
lattice theory describing the kinetic behavior
of trapping reactions
$A + B \to B$, in which both the $A$ and $B$ particles
perform  an independent
stochastic motion on a regular hypercubic
lattice.
Upon an encounter of an $A$ particle with any of the $B$ particles,
$A$ is annihilated with a finite probability; finite reaction rate
is taken into account by introducing a set of two-state
random variables - "gates", imposed on each $B$ particle, such
that an open (closed) gate
corresponds to a reactive (passive) state.
We evaluate here
a formal expression describing the
time evolution of the $A$ particle survival probability, which
generalizes our previous
results.
We prove
that
for
quite a general class of random motion of
 the species involved in the reaction process, for infinite or finite number 
of traps, and for any time $t$,
 the $A$ particle survival probability is always larger
in case when $A$ stays immobile, than
in situations when it moves.
{\flushleft PACS numbers: 05.40.-a, 02.50Ey, 82.20.-w \hspace*{\fill}}
}
\normalsize
}
\maketitle

\section{introduction}

Kinetics of chemical reactions involving diffusive
species have attracted a great deal of scientific interest
since the pioneering work by Smoluchowski \cite{smo}. Since then,
many novel and
conceptually important results have been obtained
\cite{rice,calef,blu,mi,mi2}.
In particular, it has
been proved in specific cases that the classical,
mean field chemical
kinetics does not apply, at least in low
dimensional systems
\cite{blu,mi,mi2,red,bur,bra,sza,ben}.

Trapping  $A + B \to B$ 
 reactions (TR),
involving randomly moving
$A$ and $B$ particles which 
react "when they meet" at a certain distance $R$, 
provide
an example of chemical reactions 
showing a pronounced deviation from the 
text-book predictions. 

For the TR  two 
situations were most thoroughly studied:
the case 
when $A$s diffuse while $B$s are static, 
and the situation in which the
$A$s are immobile while $B$s diffuse - 
the so-called target annihilation problem (TAP).
In the case
of static,  randomly 
placed (with mean density $b$)
traps the $A$ 
particle survival probability $P_A(t)$ shows
a non-trivial, 
fluctuation-induced behavior 
\cite{bur,bra,bal,don,pastur,gp,kh,3}
\begin{equation}
\label{traps}
\ln P_A(t) \sim - b^{2/(d + 2)} (D_A t)^{d/(d+2)}, \;\;\; t \to \infty,
\end{equation}
which is intimately related 
to many 
fundamenal problems of statistical physics 
\cite{bur,bra,bal,don,pastur,gp,kh,3,sosiska}.

Survival probability 
$P_{target}(t)$ of an immobile target $A$ of radius $R$ 
in presence of 
point-like 
diffusive traps $B$  (TAP) can be calculated exactly for any $d$
(see Refs.\cite{tach} and \cite{blu,bur,blum,szabo}):
\begin{equation}
\label{k}
P_{target}(t) = \exp\Big( - b \phi_R^{(d)}(t)\Big),
\end{equation}
where $\phi_R^{(d)}(t)$ obeys
\begin{equation}
\label{Smol}
\phi_R^{(d)}(t) = \int_0^t d\tau K_S(\tau) \sim \displaystyle \left\{\begin{array}{lll}
\displaystyle 4 \sqrt{D t/\pi},  \;\;\;   \mbox{d = 1}, \nonumber\\
\displaystyle \frac{4 \pi D t}{\ln(4 D t/R^2)},  \;\;\;   \mbox{d = 2}, \nonumber\\
\displaystyle 4 \pi D R t,    \;\;\;    \mbox{d = 3},
\end{array}
\right.
\end{equation}
where $D = D_B$ and $K_S(\tau)$ is the $d$-dimensional Smoluchowski-type "constant",
defined as the flux of diffusive particles through 
the surface of an 
immobile sphere of radius $R$. 
Decay forms in
systems with hard-core interactions between 
$B$s \cite{core} 
or with fluctuating chemical 
activity \cite{fluct} 
have also been discussed. 

On contrary, the physically most important 
case of TR when both 
$A$s and $B$s diffuse
was not solved exactly. 
It has been proven \cite{bra} 
that here $P_A(t)$
obeys
\begin{equation}
\label{general}
\ln P_A(t) = - \lambda_d(D_A,D_B) \times \displaystyle \left\{\begin{array}{lll}
\displaystyle t^{1/2},  \;\;\;   \mbox{d = 1}, \nonumber\\
\displaystyle \frac{t}{\ln(t)},  \;\;\;   \mbox{d = 2}, \nonumber\\
\displaystyle t,    \;\;\;    \mbox{d = 3},
\end{array}
\right.
\end{equation}
which equation
defines its time-dependence exactly. On the other hand, 
the factor
$\lambda_d(D_A,D_B)$ remained as yet 
an unknown function of 
the particles' diffusivities and $d$. 
Since  the time-dependence of the 
function on the rhs of 
Eq.(\ref{general}) 
follows 
precisely the 
behavior of $\int^t d\tau K_S(\tau)$, 
one might  
expect that the 
SA provides 
quite 
an accurate description for this situation and 
following its spirit 
to set $D_A = 0$ supposing that 
traps diffuse with the diffusion 
coefficient $D = D_B + D_A$.  
As a matter of fact, it has been 
often tacitly assumed
that when 
both of species diffuse $P_A(t)$ obeys 
Eq.(\ref{k}) with 
$ \phi_R^{(d)}(t)$ defined by Eq.(\ref{Smol}) and $D = D_A + D_B$.  
On the other hand, 
it has been shown that $\lambda_d(D_A,D_B)$ is 
less than the corresponding prefactor 
in $K_S(t)$  \cite{burl} and 
that  it 
may be bounded by a non-analytic 
function of $D_A$ and $D_B$
\cite{bere}.
A perturbative approach for calculation
of $\lambda_d(D_A,D_B)$, as well as  
corrections to
the SA in 1D systems were presented \cite{szabo}. 
It has been also noticed 
that $\lambda_d(D_A,D_B)$ is not 
a function of 
$D = D_A + D_B$ only, since the diffusion-reaction equation 
are not separable \cite{szabo}.   
This lack of  knowledge of the precise form  
of $\lambda_d(D_A,D_B)$, of course, constitutes
an annoying gap in the general 
understanding of the 
fluctuation phenomena in chemical kinetics. 

Recently, some very
interesting and unexpected results
have been established for
trapping $A + B \to B$ reactions
involving randomly moving species \cite{bly,benbis}, which have resolved, at least in part,
this problem.
It has been
shown that in one or two dimensions \cite{bly}, or more generally
in systems, in which the fractal dimension of the $B$ particle
trajectories is greater than the dimension $d$ of the
embedding space \cite{benbis}, (i.e. in case of the
so-called
"compact exploration" \cite{pgg}),
the leading at long times kinetic behavior of perfect
trapping  is
essentially independent of
the $A$ particle diffusion coefficient.
In other words, it has been shown \cite{bly} that 
in such low dimensions,
the leading long-time decay of the $A$ particle
survival probability in systems in which
the $A$ particle diffuses and
the decay in systems in which it 
is fixed at the origin are exactly the same.

The
derivation of this rather surprising result relies heavily
on the assumption that the
$A$ particle
has a larger probability to survive until a given
time $t$
(at least when $t \to \infty$)
if it stays
immobile rather than when it moves
randomly. In Ref.\cite{bly} some arguments have been proposed
in favor of this conjecture, based on the analysis of
the decay exponents in systems with a finite number of $B$s. Subsequently,
it was shown rigorously in Ref.\cite{blybis} that it is indeed the case
in a one-dimensional continuum; on the other hand, 
the claim that Ref.\cite{blybis} presents a rigorous proof
of this conjecture for $d = 2$ does not seem to be justified;
as a matter of fact, Eq.(2) of Ref.\cite{blybis} does not make sense
for $d = 2$ and the logarithmic correction does not follow
from it, since particle's radius is not taken explicitly into account
within the approach of Ref.\cite{blybis}.

On the other hand, in a
recent article \cite{mor},
this conjecture has been examined
rigorously within the context of reactions
between the particles executing
random walks on $d$-dimensional lattices;
here, such a conjecture has been
referred to as the
"Pascal principle",
since it is reminiscent of a famous philosophical assertion
of Blaise Pascal, who claimed that
"all
misfortune of man comes from the fact that
he does not stay peacefully in his room" \cite{pas}.
In Ref.\cite{mor},  we
showed that as $t \to \infty$
the Pascal principle-like inequality
between the survival probabilities
of a diffusive and of an immobile $A$ particles is valid in any dimension,
provided that the
$A$ particle performs some rather
general continuous-time jump process on a
hypercubic lattice, while the
$B$ particles perform independently
a
discrete time lattice jump process, which also
satisfies some rather
natural assumptions. The same conclusion
was also obtained in Ref.\cite{mor}
for a much more general
case of
stochastically-gated
reactions,
which mimic situations with finite
elementary reaction act constants.

We also emphasize
that very similar Pascal principle-like inequality
 has been proven earlier
for the process of
hopping transport
of an excitation
on a disordered array of
immobile
donor centers in presence of
randomly placed, immobile
quenchers \cite{burl}.  We note, as well,  that
recent results obtained for
the  ballistic
$A + A \to 0$ annihilation
process \cite{jaroslaw} are compatible
with such a principle.

On the other hand, the analysis in Ref.\cite{mor} is rather
condensed and moreover, some of the assumptions invoked, as well
as some of the constraints imposed on particles' random walks,
seem to be unnecessary and thus can be safely relaxed.
Consequently, our purpose here is to complete the proof of the
Pascal principle-like inequality between the survival
probabilities of the diffusive and immobile $A$ particles and to
extend it in several directions. In particular, we proceed to show
that the Pascal principle-like inequality holds at any finite time
$t$, as well as for both infinite and finite number of traps. 
Moreover, we shall consider here the case when the
chemical activity of the $B$ particles fluctuates in time
 between active and inactive states. We set out to
show that
the
Pascal principle also applies for this much more complex and realistic situation.
 As in our previous
work \cite{mor}, we will focus here solely on the  lattice
formulation
of the model.
The continuous-space case, which requires much more delicate analysis,
will be studied elsewhere.

The paper is structured as follows: In Section 2 we formulate the model,
introduce basic notations and define the properties of reaction and random walks
executed by the species involved.
In section 3 we will focus on the reaction kinetics
in case of perfect trapping; that is,
on the case when the $A$ particle gets annihilated
with probability $1$
upon the first encounter with any of the traps $B$.
Further on,  in section
4,
the Pascal principle-like inequality
between the survival probabilities of a diffusive and immobile $A$ particles
will be extended to the
case when the chemical activity of the $B$
particle
fluctuates randomly between active and inactive states,
which mimics
more realistic
situations in which
an annihilation of an $A$ particle upon its encounter with any of the $B$s
takes place with a finite probability.
We consider here a rather general case when such activity
fluctuations can be correlated in
time.
Some intermediate calculations, as well as analysis of the behavior in some special
cases are relegated to the Appendices A and B, in which, in particular,
the special case of Polya random walks is considered.

\section{Lattice model of trapping reactions between mobile species}

It is well-known that lattice models of
diffusion-controlled reactions yield, at least for sufficiently large times,
the kinetic laws that are
essentially the same as those obtained within
the continuous-space
descriptions. Thus in the present
work we shall consider a
\textit{lattice model} of trapping reactions,
which will simplify significantly
our analysis. One of advantages of such a consideration, apart of the
fact that it allows for much more lucid analysis than in
the continuous-space limit,
is that we are not forced to attribute to particles a finite, non-zero
 radius, which allows to consider the behavior in systems of \textit{any}
spatial dimension.
Finally,
for sake of simplicity,
we will restrict our analysis here
to hypercubic lattices; most of
the
results, of course, could be readily
extended to other types of embedding
lattices.

Consider $N_A$ particles of type
$A$ and $N_B$ particles of type
$B$, which are initially placed at
random at the sites
of a $d$-dimensional hypercubic lattice, containing $M$ sites.
All  particles perform  independent jump processes on the nodes of
the lattice. Each $A$ particle can be destroyed (in the general case with a finite
probability) as soon as this $A$ particle appears on the same lattice site
simultaneously with
any of the $B$ particles. The $B$ particle remains unchanged
after the reaction event,
which corresponds to the annihilation mechanism
\begin{eqnarray}
\displaystyle
 A + B \to B,
\end{eqnarray}
and represents the customary trapping-like reaction.
One can also envisage a more general catalytic
reaction process of the form:
\begin{eqnarray}
\displaystyle
\eqnum{5'} A + B \to C + B
\end{eqnarray}
where the catalyst $B$
promotes the transformation of an
$A$ into some product molecule $C$,
the product molecule $C$ being immediately
extracted from the system.

In regard to the reaction probability, we will distinguish between two situations:
the one of perfect trapping, or purely diffusion-controlled trapping,
in which case any $A$ gets annihilated with probability
$1$ upon the first encounter with any of $B$s, and that of imperfect
trapping for which the annihilation of the $A$ by any $B$ takes place with a finite
probability $< 1$. To mimic this condition, we will introduce a set of additional
random variables, attached to each $B$ particle, which will describe their
instantaneous reactive activity. Finally,
we will assume in what follows
that collisions (simultaneous encounters)
between two (or more)
$A$ particles are possible and
do not affect these $A$ particles or their random walks,
 and similarly, that
collisions between
the $B$ particles are possible and do not lead to any reactions.
In other words,
neither $A$ nor $B$ particles have hard-core interactions
and no single-species
reactions may take place.

Now, as far as particle motions are concerned,
 we face here the
following problem: on one hand,
in regard to dynamics of $A$ and $B$ particles,
we have to
define
two different random processes
with different characteristics, e.g. diffusion coefficients,
which may be used afterwards as tunable parameters.
On the other hand,
these random processes must allow a rigorous
analysis, which is not always the case.
If we choose, for example, that both species perform
random hopping motion in discrete time, then it will be quite difficult
to work out a rigorous formalism
in which two random
processes have different "waiting" times at lattice sites.
If, on contrary, we choose that both processes
evolve in continuous time, then we will
face purely mathematical difficulties in
treatment of the events
 in which the particles of different species
appear simultaneously
at the same lattice site. Not forsaking the generality,
we thus choose here the "mixed" case, which seems to us
most suitable for the rigorous description.

We thus assume that all $A$ particles perform
identical and independent continuous-time jump processes,
so that
they can jump at any time moment
from one lattice site to any other site. No
other hypothesis or additional constraints on their motions are required.

Further on, we suppose that all $B$s perform identical
and independent discrete-time random walks; that is,
at any integer time $n\geq0$,
any $B$ particle can jump from a lattice site $y$ to site $y'$ with a
given probability $p(y'| y)$, where $y'$ can be identical to $y$, i.e.
$B$ can remain at the site which it occupies at time moment $n$.
Let  $Y_n$ denote the
position of a given $B$ particle
at time moment
$n\geq0$.
We assume that this random
walk satisfies the following conditions:

\textbf{\textit{(i)}}  the random walk is homogeneous in space and in time:
the probability $p(y'|y)$ is independent of time and of the initial
position, such that
\begin{eqnarray}
\displaystyle
p(y'|y)=p(y'-y).
\end{eqnarray}

\textbf{\textit{(ii)}} the uniform distribution $p(y) = 1/ M$ is
stationary for the $B$ particles. We assume that it holds at time
0, and hence, it is realized at all times. This condition implies
the bilateral normalization relation
\begin{eqnarray}
\displaystyle
\sum_yp(y'|y)=1.
\end{eqnarray}
We remark that
a stronger condition
would be to assume that
the probability $p(y'| y)$ satisfies the detailed balance, i. e.
\begin{eqnarray}
\displaystyle
\nonumber p(y'|y)=p(y|y')
\end{eqnarray}
which, together with the condition in eq.(2),
implies that
\begin{eqnarray}
\displaystyle
p(y-y')=p(y'-y).
\end{eqnarray}
However, the condition in eq.(3),
which clearly follows from eq.(4),
is sufficient for obtaining our main results.

\textbf{\textit{(iii)}} at any time $n$, the conditional probability
$P(Y_n = y|Y_0 = 0)$ of finding a given $B$ particle
 at an arbitrary position $y$ at time moment
$n$, provided that it started its random walk at the origin,
does not exceed  the return probability
$P(Y_n = 0 |Y_0 = 0)$. That is,

\begin{eqnarray}
\displaystyle
P(Y_n = y|Y_0 = 0)\leq P(Y_n = 0 |Y_0 = 0)\equiv R_n
\end{eqnarray}

This condition seems to be quite plausible
for any symmetric random motion
in a uniform medium
if eq.(4) holds,
since here $P(Y_n = y |y_0 = 0)$ is invariant upon reversal $y$
into $-y$. Hence,
this probability should always have
 an extremum for $y = 0$,
which is likely to be a maximum. It should be noticed, however,
that inequality in eq.(5) does not hold exactly for the usual
Polya random walk, when the particle jumps at one of neighboring
sites at each integer time moment. For instance, on a one
dimensional lattice of unit spacing, the inequality in eq.(5) is
not satisfied if $n$ and $y$ are odd, since in this case the
return probability $R_n \equiv 0$. Nevertheless, one easily
obtains (without any significant lack of generality) random walks
satisfying condition \textbf{\textit{(iii)}}: for instance, one
may consider a modified Polya random walk on a $d$-dimensionnal
hypercubic lattice, such that at each integer time moment   a
walker has  a probability $p_0$ to remain at the site it occupies,
and a probability $(1-p_0)/2d$ to jump at one of the neighboring
sites. In this case, it can be shown that the inequality in eq.(3)
is verified if $p_0 \geq 1/2$. More generally, eq.(5) holds for
any homogeneous and symmetric probability $p(x|y)$ if $p(x|x)
\equiv  p(0) \geq 1/2$ (see Appendix A). Furthermore, our
conclusions can be extended to cover the case of the Polya random
walks, as shown in Appendix B, where this special question is
discussed in detail.

We finally remark that in the continuous-space case, in which the
random walk is replaced by a Brownian motion, such a question does
not arise at all, since here the probability density is always
maximal and centered around the initial position. Thus the
inequality in eq.(5)  appears very naturally in unbiased diffusion
problems, but it can also be verified for non symmetric jump
probabilities.

We close this section by adopting some conventions on how
to introduce reaction events into the model.
We assume that a given $B$ particle can only annihilate $A$ at integer
times $n>0$. If at a non integer time $A$ jumps on a site which is occupied
by a particle $B$, it will be only annihilated
at the next integer time
$n$.
Finally, we remark that the probability that an $A$ particle performs a
jump exactly at an integer time is 0, which allows to neglect
consideration of such events.
Note also that all these assumptions won't change the global behavior
of the system. They thus merely serve for convenience of exposition.

\section{Perfect trapping on a lattice.}

We
consider first the case of perfect trapping in which case an annihilation of an $A$
particle takes place at the first encounter with any of $B$ particles.
Our aim here is to demonstrate, in a rigorous
way, the Pascal principle-like assertion that the survival probability of an  $A$
particle which moves randomly on a lattice is less or equal to the survival
probability of  an immobile $A$ particle.

\subsection{Mean-field kinetics of the trapping reaction.}

We start with a reminder on the predictions of a conventional
mean-field approach [2].
One notices first that,  clearly,
the average number $<N_A(n)>$ of
$A$ particles surviving up to an integer time $n$
is the sum of
probabilities that a given $A$ particle survives up to this time moment
$n$.
Since all of them have identical evolution laws, one has
\begin{eqnarray}
\displaystyle
\nonumber \left\langle N_A (n) \right\rangle =  N_A(0) \Psi(n),
\end{eqnarray}
where  $\Psi(n)$ denotes
the survival probability of a single particle
$A$.
Since the particles $B$ are completely insensitive
(as far as their
motions are concerned) to particles $A$,
$\Psi(n)$ can be evaluated
independently for
each particle $A$.
Thus, it is legitimate to consider only
the survival of a single
$A$ particle in presence of $N$ particles $B$.

In terms of the conventional mean-field kinetics [2], one obtains
then an exponential decay
form for $\Psi(n)$:
\begin{eqnarray}
\label{exp}
\displaystyle
\Psi(n)=\exp{(-kbn)},
\end{eqnarray}
which should hold in any dimension $d$. In the last equation
$k$ is the reaction constant and $b$ stands for the mean
density of the $B$ particles.

Note that in case of perfect trapping eq.(\ref{exp})
becomes senseless, since here $k = \infty$.
Indeed, it has been well-known for a long time,
both for the continuous-space and
lattice models, that the decay law in eq.(\ref{exp})
does not hold,
at least for $d = 1$ and $d = 2$
\cite{smo,blu,red,bur,bra,sza,ben},
so that
mean-field approach fails and
a detailed stochastic theory is needed.

To illustrate the deviations from the mean-field behavior in eq.(\ref{exp})
and the actual decay forms,
let us consider
the case when
one of the species only is moving
\cite{red,bur}. In this illustration, we  follow closely
the methods outlined in
Ref.\cite{benter}.

\subsection{Survival probability of an $A$ particle.}

Let us call  $\Gamma_A$  the $A$ particle trajectory, and first
suppose that it is given. Then,   we denote as $x_0 = 0,
x_1,\ldots,x_n$  the successive positions of the $A$ particle at
the integer times $t_0=0, t_1,\ldots,t_n$.

We suppose next that the waiting time of $A$ at each lattice site,
i.e. the time which an $A$ particle spends on this site between
successive hops, is a stochastic variable, so that two successive
positions are not necessarily different, and not necessarily
nearest neighbors.

Further on,
we denote the $i$th $B$ particle, $i = 1, \ldots, N$, as $B_i$
and as $\Gamma_{B_i}$ -
the stochastic trajectory of
this particle. Next,
let $Q_i(n|\Gamma_A)$ be
\textit{the conditional probability that $B_i$ does not destroy
$A$ up to time moment $n$},
for a given trajectory $\Gamma_A$ of $A$.
Because all $B_i$s move and act independently of each other,
the conditional
probability $\Psi(n|\Gamma_A)$ that the
particle $A$ survives up to time moment
$n$ for a given $\Gamma_A$, factorizes
\begin{eqnarray}
\displaystyle
\Psi(n|\Gamma_A) = \prod_{i=1}^N Q_i (n|\Gamma_A)
\end{eqnarray}
and hence,
the overall
$A$ particle survival probability obeys
\begin{eqnarray}
\displaystyle
\Psi(n) = \langle \Psi(n|\Gamma_A) \rangle_{\Gamma_A},
\end{eqnarray}
the average being taken over all possible trajectories of
$A$ from $t = 0$  to  $t = n$.
Furthermore, since all $B$ particles
are
identical, one has that
 $Q_i(n|\Gamma_A) = Q(n|\Gamma_A)$ for all $i$, and
hence
\begin{eqnarray}
\displaystyle
\Psi(n) = \langle Q(n|\Gamma_A)^N  \rangle_{\Gamma_A}
\end{eqnarray}
where, once again, the average is being
taken over all possible trajectories
$\Gamma_A$ of the $A$ particle.

\subsection{The survival probability in the thermodynamic limit.}

Let us denote
$Y_0, Y_1, \ldots, Y_n$ the successive positions of a given $B$
particle at time moments $0, 1, \ldots, n$,
and  $\Gamma_{y_0}$ - a trajectory starting from  $Y_0=y_0$ at time $0$. One
can write then
\begin{eqnarray}
\label{1}
\displaystyle
 Q(n|\Gamma_A) & = & \langle Q(n|\Gamma_A,y_0) \rangle_{y_0},
\end{eqnarray}
where
 $Q(n|\Gamma_A,y_0)$  stands for
the conditional probability that a given $B$ particle,
starting its random walk from position $y_0$ at time moment $0$,
does not destroy $A$ until time moment
$n$ for a given trajectory $\Gamma_A$.
The brackets  $<  \ldots >_{y_0}$  in eq.(\ref{1})
denote averaging with respect to all
possible initial positions $y_0$ of a given $B$ particle.

We now assume that the probability of
the initial position $Y_0$ is uniformly distributed among the
$M$ available sites. Then we have
\begin{eqnarray}
\displaystyle
\nonumber \langle Q(n|\Gamma_A,y_0) \rangle_{y_0} & = & \frac{1}{M} \sum_{y_0} Q(n|\Gamma_A,y_0)
\end{eqnarray}
and eq.(9) can be written
\begin{eqnarray}
\displaystyle
\Psi(n) & = & \langle \{ 1 - \frac{1}{M} \sum_{y_0} \left( 1-Q(n|\Gamma_A,y_0) \right) \}^N  \rangle_{\Gamma_A}
\end{eqnarray}

Turning next to
 the thermodynamic limit,
i.e. setting
$N \to \infty$ and $M \to \infty$, while
keeping their ratio fixed, $N/M \to b$,
$b$ being the concentration of the $B$ particles,
one obtains for the $A$ particle
survival probability at time $n$ the following expression
\begin{eqnarray}
\displaystyle
\Psi(n) & = & \langle \exp{\{-b\sum_{y_0} \left( 1-Q(n|\Gamma_A,y_0) \right)\}} \rangle_{\Gamma_A}
\end{eqnarray}
Hence,
the survival probability $\Psi(n)$ is simply
related to the probability that a given
$B$, starting from $y_0$,
destroys $A$ at some time $t \leq n$, for a given trajectory of $A$,
which is
\begin{eqnarray}
\displaystyle
P(n|\Gamma_A,y_0) & = &  1 -  Q(n|\Gamma_A,y_0)
\end{eqnarray}
Similar results were obtained \cite{blu,red,bur,bra,sza,ben,benbis,benter}
in the particular case when the $A$ particle
is immobile, i.e. for the so-called target annihilation problem.
In this particular case there
 is no averaging
over $\Gamma_A$ as in the previous
formulas, and the integral reaction rate is thus defined by
\begin{eqnarray}
\displaystyle
K(n|\Gamma_A) \equiv \sum_{y_0}P(n|\Gamma_A,y_0),
\end{eqnarray}
which replaces in this case
the term  $k n$ of the conventional
 kinetic law in eq.(6).
On
contrary, in more realistic situations when
 $A$ also moves,
the average over the trajectories $\Gamma_A$ makes the
explicit calculation of the survival probability impossible in most
cases.

\subsection{A basic inequality.}

Let us define  $P^1(k|\Gamma_A,y_0)$
as \textit{the conditional probability that} $B$, starting from $y_0$ at time $0$,
 \textit{meets $A$  for the first time} at time $k$, given
the trajectory $\Gamma_A$. Then, the conditional probability
$P(n|\Gamma_A,y_0)$  that $B$, starting from $y_0$,
destroys $A$ at or before time moment
$n$ is given by
\begin{eqnarray}
\displaystyle
P(n|\Gamma_A,y_0) =  \sum_{0< k\leq n} P^1(k|\Gamma_A,y_0)
\end{eqnarray}
The conditional probability that the trajectory of $B$ (extended after
the possible annihilation of $A$)
meets $\Gamma_A$ at time $n$ (not
necessarily for the first time) satisfies the equation
\begin{eqnarray}
\displaystyle
\nonumber P(Y_n = x_n| Y_0 = y_0) = \\
\sum_{0\leq k<n} P(Y_n = x_n| Y_k = x_k)P^1(k|\Gamma_A,y_0),
\end{eqnarray}
where $P(Y_n = x_n| Y_k = x_k)=\delta_{x_n,x_k}$ and  $P^1(0|\Gamma_A,y_0)=0$.

Summing both sides of
the last equation over all initial positions $y_0$ and using the
relation in eq.(3), which applies to  $P(Y_n = x_n|Y_0=y_0)$, we obtain
\begin{eqnarray}
\displaystyle
1=\sum_{0\leq k<n}P(Y_n = x_n| Y_k = x_k)S(k|\Gamma_A),
\end{eqnarray}
where we have used the notation
\begin{eqnarray}
\displaystyle
S(n|\Gamma_A) =  \sum_{y_0} P^1(n|\Gamma_A,y_0)
\end{eqnarray}
Next, using the inequality in eq.(5) we obtain from  eqs.(15) and (18), the following
basic inequality
\begin{eqnarray}
\displaystyle
1\leq\sum_{0\leq k<n}R_{n-k}S(k|\Gamma_A),
\end{eqnarray}
where $R_{n-k}$ is the probability of return to the starting point in
$n-k$ steps, which is
a well-known quantity for all classical random walks.

We notice that if the $A$ particle stays immobile,
the inequality in eq.(19) becomes
the \textit{equality}, since here
$x_k=0$ for all times $k$.

Now, let $\widehat{F}(s)$
denote the generating function
 of some function $F(n)$,
\begin{eqnarray}
\displaystyle
\widehat{F}(s) = \sum_{n>0}F(n) s^n
\end{eqnarray}
Multiplying both sides of the inequality in eq.(19) by $s^n$ and performing summations,
we have then
\begin{eqnarray}
\displaystyle
\left( \frac{1}{1-s} \right)  \leq  \widehat{R}(s) \widehat{S}(s|\Gamma_A),
\end{eqnarray}
where $\widehat{R}(s)$ is the generating function of the
return probability $R_n$, while
\begin{equation}
\widehat{S}(s|\Gamma_A) = \sum_{n>0} \left( \sum_{y_0} P^1(n|\Gamma_A,y_0) \right) s^n
\end{equation}
Note that again,
the inequality
in eq.(21) becomes the equality in the particular case
when
$A$ is immobile, so that
\begin{eqnarray}
\displaystyle
\widehat{S}(s|0) \leq \widehat{S}(s|\Gamma_A)
\end{eqnarray}
where $S(n|0)$ denotes the $S(n |\Gamma_A)$ in case when $A$ is
immobile.

On the other hand, one readily
notices from eqs.(12) to (14) and
(18),  that the $A$ survival probability at time $n$ is just
\begin{eqnarray}
\displaystyle
\Psi(n) = \langle\exp{(-bK(n|\Gamma_A))}\rangle _{\Gamma_A}
\end{eqnarray}
where
\begin{eqnarray}
\displaystyle
K(n|\Gamma_A) = \sum_{0\leq k \leq n}S(k|\Gamma_A)
\end{eqnarray}

In the limit $n\to \infty$ this expression coincides formally
with the
generating function of $S$,  if $s \to 1$,
which suggests that inequality
(23) corresponds, at least asymptotically, to the similar
inequality
\begin{eqnarray}
\displaystyle
K(n|0) \leq K(n|\Gamma_A)
\end{eqnarray}
In this inequality
the right-hand side
corresponds
to the case of an
immobile particle $A$.
Consequently, the inequality in eq.(26)
implies that the annihilation is faster if $A$ moves than
if  it is immobile,
in agreement with the Pascal principle.

However, the generating functions $\widehat{S}(s|0)$ and
$\widehat{S}(s|\Gamma_A)$ tend to $\infty$ when $s\to 1$,
and the derivation of eq.(26) requires
a more careful analysis, which is the purpose of the next paragraph.

\subsection{General form of Pascal principle.}

Let us turn back to the inequality in eq.(19) and recall that
it becomes an equality in the case when the $A$ particle does not move. Then, we may
formally rewrite
the inequality in eq.(19) in the following form:
\begin{eqnarray}
\displaystyle
0\leq \sum_{0\leq k \leq n}R_{n-k}[S(k|\Gamma_A)-S(k|0)]
\end{eqnarray}
Next, let us introduce two auxiliary functions
$L_n$ and $M_n$, such that
\begin{eqnarray}
\displaystyle
L_n = S(k|\Gamma_A)-S(k|0)
\end{eqnarray}
and
\begin{eqnarray}
\displaystyle
M_n = \sum_{0\leq k \leq n}L_k = K(n|\Gamma_A)-K(n|0)
\end{eqnarray}
By definition, we have $R_0 = 1$ and  $L_0 = 0$.
Then, the inequality in eq.(31)
can be straightforwardly written as
\begin{eqnarray}
\displaystyle
M_n \geq \sum_{1\leq k \leq n-1}\left( R_{n-1-k}-R_{n-k} \right) M_k
\end{eqnarray}
Now, it can be shown (see Appendix A)
that  $R_k$ is a decreasing function
of $k$.
Asuming that  it has been proved that  $M_k \geq 0$  for all  $0
\leq  k \leq  n - 1$, it follows from eq.(34) that  $M_n \geq  0$,  and inequality
in eq.(31)
is proved by induction, since $M_0 = 0$.

Consequently, for any time $n$,
the $A$ particle survival
probability $\Psi(n)$, defined by eq.(24),
in case when the $A$ particle does not move
is less or equal to the
survival probability in case
 when $A$ is mobile.
This result is much stronger than the asymptotic form of the
Pascal principle obtained in Ref.\cite{mor}.

We note  also that this result 
holds for any finite system with a finite number
of traps (i.e. not necessarily in the thermodynamic limit).
Here, the survival probability is given by eq.(9), in which equation
$Q(n|\Gamma_A)$ denotes the probability that
a given $B$ particle
does not meet
$A$ before or at time moment $n$.
In fact, the probability  $P^1(n|\Gamma_A)$
that a given $B$ particle meets $A$ for the first time at time moment
$n$ obeys
\begin{eqnarray}
\displaystyle
P^1(n|\Gamma_A) = Q(n-1|\Gamma_A)-Q(n|\Gamma_A)
\end{eqnarray}
and we have
\begin{eqnarray}
\displaystyle
Q(n|\Gamma_A)=1-\sum_{0\leq k \leq n}P^1(k|\Gamma_A)
\end{eqnarray}

In Section \S3.3.  we introduced $P^1(k|\Gamma_A)$ as \textit{the conditional
probability that $B$}, starting from $y_0$ at time 0, \textit{meets
$A$  for the first time} at time $k$. Particles $B$ are uniformly
distributed among the $M$ lattice sites at time 0 (and at all time as
well), so that, with the notations used in eq.(18), we find that
\begin{eqnarray}
\displaystyle
P^1(k|\Gamma_A) = \frac{1}{M}\sum_{y_0}P^1(k|\Gamma_A,y_0) = \frac{1}{M}S(n|\Gamma_A),
\end{eqnarray}
while eq.(35) reads
\begin{eqnarray}
\displaystyle
Q(n|\Gamma_A) = 1-\frac{1}{M}K(n|\Gamma_A)
\end{eqnarray}
Thus, the inequality in eq.(26)
implies that, whatever may be the number of $B$ particles,
particle $A$ has a higher probability to escape if it stays immobile, than if it moves.
Note that this conclusion had been
drawn previously by Bray and Blythe \cite{bly} for systems with
 a
finite number of traps within the context of survival of a mobile
pray $A$ in presence of a finite
number of predators $B$. Within this context, the Pascal principle-like
inequality in eq.(26) appears to be even more sound.

To close this section we note that
the inequality in eq.(26)
may be questioned
for
usual Polya random walks \cite{hug},
since the condition in eq.(5) is not strictly fulfilled.
It is shown in Appendix B how our results can be extended to this case.

\section{Imperfect trapping.}

\subsection{Time correlated chemical reactivity fluctuations}

We now modify the model presented in Section 2, assuming that
the $A$ particle has
 a finite probability
(which may depends on time) to survive when encountered by a $B$
particle. This case occurs if the reaction is not purely
controlled by diffusion: at each encounter, another stochastic process
arises and allows  the reaction to be eventually completed, or to
fail.
This process is an elementary reaction act.

If there is a single $A$ particle, it is physically plausible to assume that
at each of its encounters with any of the $B$s, the latter
 can be either in a passive internal
state with a (possibly time-dependent) probability $p(t)$
($(0<p(t)<1)$), or in an active state with probability $1 - p(t)$.
In the latter case, the $A$ is destroyed, whereas it remains
intact if $B$ is passive and they may harmlessly coexist until the
$B$ changes its reactive state. In Ref.\cite{mor} we have already
addressed this problem, assuming that this reaction
 probability was constant and independent of
all prior events. In many
circumstances, however, this assumption is not justified, and, in particular,
the survival probability of the  $A$ particle
during its encounter with any of the $B$s
may itself depend on the trajectory of these
particles. We will not treat
this difficult problem in general, but
only consider a special situation, in which the $A$ particle survival probability
depends on some internal, stochastic property of particle
$B$. Then it is possible to take into account the time correlations of
its fluctuations. This model can be justified as an approximation of
certain phenomena, such as possible fluctuations in the activity of
the catalyst in a chemical reaction \cite{morbis}.

More precisely, we assume \cite{benter,morter} that each particle $B$
can be in a passive state 0 or in an activated state 1,
the waiting time $T_i$ in state $i$ being a stochastic variable
independent of prior events, distributed
following an exponential law
\begin{eqnarray}
\displaystyle
P(T_i>t)=\exp{(-\lambda_it)} \mbox{ (i=0 or 1)},
\end{eqnarray}
where $\lambda_0$ and $\lambda_1$ are given positive constants.

Now, the  transition probability
for the internal state $I(t)$ of $B$ is
then given by the well-known ``random telegraph'' law \cite{fel}
\begin{eqnarray}
\displaystyle
P(I(t)=i|I(0)=j) = \alpha_i + (\delta_{ij}-\alpha_i)e^{-\lambda t}
\end{eqnarray}
with
\begin{eqnarray}
\displaystyle
\lambda = \lambda_0 + \lambda_1 \mbox{ and } \alpha_0=\lambda_1/\lambda, \mbox{ } \alpha_1=\lambda_0/\lambda
\end{eqnarray}

Thus, $\alpha_0\equiv p$ is the asymptotic probability that $A$
survives a collision with $B$, whereas $\alpha_1=1-p$ is the
asymptotic reaction probability at each encounter. The elementary
reaction act constant $k$, mentioned in the beginning of Section
2, is just $k \sim (1-p)/p$. We logically assume that the internal
state of $B$ is initially in its stationary probability
distribution, as well as at the first encounter of with $A$, but
at the next encounter the law given by eq.(40) should be used.

Extending eq.(16) to the present situation with a stochastic
elementary reaction act, we find
\begin{eqnarray}
\displaystyle
\nonumber &\alpha_1& P(Y_n = x_n|Y_0 = y_0) =
\underline{P}^1(n|\Gamma_A,y_o) + \\
\nonumber &+& \sum_{0\leq k\leq n-1}(\alpha_1+\alpha_0e^{-\lambda (n-k)}) \times\\
&\times& P(Y_n = x_n|Y_k = x_k) \underline{P}^1(k|\Gamma_A,y_o)
\end{eqnarray}

In fact,  $\alpha_1P(Y_n = x_n|Y_0 = y_0)$ is the probability that
$B$ meets $A$ at time $n$ while it is in its active state: the
probability for $B$ to be in its active state at time $n$ is
independent of the trajectories of $B$ or $A$, and is given by the
stationary value $\alpha_1$, since no value of the internal state
is assigned before time $n$.  Furthermore,
$\underline{P}^1(k|\Gamma_A,y_o)$ is the probability that $B$
meets $A$ in its active state at time $k$ \textit{for the first
time} after 0, with $\underline{P}^1(0|\Gamma_A,y_o)=0$. If  $B$
meets $A$  in its active state at time $n$, then necessarily the
same situation occurred for the first time at some time $k$,
$0<k\leq n$. If  $k<n$, then the probability for $B$ to be again
in its active state is given by eq.(40), which gives rise to the
last term in the right hand side of eq.(42).

Equation (42)
can be written in a more compact form
\begin{eqnarray}
\displaystyle
\nonumber &\alpha_1& P(Y_n = x_n|Y_0 = y_0) = \sum_{0\leq k\leq n}(\alpha_1+\alpha_0e^{-\lambda
(n-k)}) \times \\
&\times& P(Y_n = x_n|Y_k = x_k)\underline{P}^1(k|\Gamma_A,y_o)
\end{eqnarray}
Summing both sides of it over
the initial position $y_0$, we then obtain
\begin{eqnarray}
\displaystyle
\nonumber \alpha_1 =  \sum_{0\leq k\leq n}(\alpha_1+\alpha_0e^{-\lambda (n-k)}) \\
P(Y_n = x_n|Y_k = x_k)\underline{S}(k|\Gamma_A)
\end{eqnarray}
where $\underline{S}(k|\Gamma_A)$ is the probability that the annihilation of $A$
occurs at time $k$, for a given trajectory of $A$.

Using next the inequality in eq.(5), we find  the following
relation
\begin{eqnarray}
\displaystyle
1 \leq \sum_{0\leq k\leq n} \left( 1+\frac{\alpha_0}{\alpha_1}e^{-\lambda (n-k)} \right) R_{n-k}\underline{S}(k|\Gamma_A)
\end{eqnarray}

We now make use of the generating functions technique
and obtain, returning to
the notation  $p\equiv\alpha_0$, that
\begin{eqnarray}
\displaystyle
\frac{1}{1-s} \leq \left[
\widehat{R}(s)+\frac{p}{1-p}\widehat{R}(se^{-\lambda}) \right]\widehat{\underline{S}}(s|\Gamma_A)
\end{eqnarray}
which becomes  the equality in case $A$ is immobile.
The conclusions
follow as previously: the generating function of  the conditional
reaction probability at
time $n$  is minimal if $A$ is immobile, i.e.,
\begin{eqnarray}
\displaystyle
\widehat{\underline{S}}(s|0) \leq \widehat{\underline{S}}(s|\Gamma_A)
\end{eqnarray}
Consequently,
 the integral reaction rate $\underline{K}(n|\Gamma_A)$ is
minimal if $A$ stays immobile
\begin{eqnarray}
\displaystyle
\underline{K}(n|0) \leq \underline{K}(n|\Gamma_A),
\end{eqnarray}
the left hand sides of eqs.(47) and (48)
denoting the quantities corresponding to an immobile $A$, respectively.

The fact that eq.(48)
holds at any time $n$ can be proved directly by using
 the
inequality in eq.(46) exactly
in the same fashion
as it has been done in Section 3.3. (when all collisions are  reactive ($p=0$)).

\subsection{Asymptotic reaction kinetics}

Let us first consider the special case when A is immobile. Then, the
asymptotic kinetic behavior follows from eq.(46):
\begin{eqnarray}
\displaystyle
(1-s)\widehat{\underline{S}}(s|0) = \left[
\widehat{R}(s)+\frac{p}{1-p} \widehat{R}(se^{-\lambda}) \right]^{-1}
\end{eqnarray}

In one and two dimensions, $\widehat{R}(s)$ tends to infinity when $s\to 1$, so that the
terms due to the reactivity fluctuations in the right hand side of
eq.(49) do not affect the kinetics, which proceeds exactly in the same way as in the case of perfect
trapping reactions.

In three dimensions $\widehat{R}(s)$ tends to a finite limit $1/S$
when $s \to  1$, where $S$ is the probability  that  a given $B$
 particle never returns to its initial position (see Appendix A).
Then,
the left-hand-side of eq.(49)
tends to an effective, "apparent" reaction constant $\underline{k}$,
which satisfies the inverse addition relation
\begin{eqnarray}
\displaystyle
\frac{1}{\underline{k}} = \frac{1}{S} + \frac{p}{1-p}\widehat{R}(e^{-\lambda})
\end{eqnarray}
if $A$ is immobile \cite{benter,morter}.

Equation (50) shows that if $A$ is immobile, the reaction rate
$\underline{k}$ is \textit{an increasing function of the relaxation frequency}
$\lambda$ of the activity fluctuations, so that the survival probability decreases
with $\lambda$, if $p$ is maintained constant. It can be shown that this remarkable
property is more general and is also valid if both particles are mobile \cite{morter}.
In the case of an infinite relaxation frequency, or uncorrelated fluctuations,  eq.(50) becomes
\begin{eqnarray}
\displaystyle
\frac{1}{\underline{k}} = \frac{1}{S} + \frac{p}{1-p}
\end{eqnarray}

Equations (50) and (51) are particular cases of the "inverse addition
law" which is well-known in chemistry \cite{ben,morbis,col}.
In fact, such a law is valid if  the reaction can be considered as a succession of
independent steps, which is the case for uncorrelated fluctuations.
It was discussed in this context in our previous works \cite{morbis}.

We shall now partially extend these results for a \textit{mobile particle
$A$}. In fact, in one or two dimensions it has been shown \cite{bly,benbis} that when $A$ is annihilated
as soon as it meets any of $B$ particles ($p=0$), the survival probability
does not depend on the motion of $A$ in the limit $n \to \infty$
\begin{eqnarray}
\displaystyle
\nonumber \Psi(n;p=0) \approx\Psi_0(n;p=0)
\end{eqnarray}
where $\Psi_0$  is the survival probability in case of an immobile $A$.

If $A$ has a finite probability to survive at each encounter, the
overall survival probability $\Psi(n;p,\lambda)$ is obviously larger than if $p =
0$. However, we have shown in eq.(47)
 that $\Psi(n;p,\lambda)$  is smaller than the
survival probability  $\Psi_0(n;p,\lambda)$  in case of an immobile particle, and
the latter  is asymptotically independent of $p$. Thus, we can write
\begin{eqnarray}
\displaystyle
\nonumber \Psi_0(n;p=0) \approx\Psi(n;p=0) \leq \Psi(n;p,\lambda) \\
\leq \Psi_0(n;p,\lambda) \approx \Psi_0(n;p=0)
\end{eqnarray}
so that for large $n$
\begin{eqnarray}
\displaystyle
\Psi(n;p,\lambda)\approx\Psi_0(n;p,\lambda) \approx \Psi_0(n;p=0)
\end{eqnarray}
which shows that in one and two dimensions the reactivity fluctuations
of $B$ as well as the motion of $A$ do not affect the survival probability
of $A$ in the asymptotic limit $n\to \infty$, except if $B$ is immobile,
in which case the survival probability has a very different and
unusual behavior \cite{don}.

In three dimensions, on contrary,
the fluctuations actually change the
reaction kinetics. The survival probability decreases exponentially
and the overall reaction rate $\underline{k}$ is given by the inverse
addition law in eq.(50).
However, it is unclear if such a law still holds
when both particles move, since then we have only proved inequality
in eq.(53). Thus, the first and the last approximate equalities in eq.(51) do
not hold in three dimensions, whereas the relations in eq.(51) are
valid. It is  known \cite{bra} that for large $n$, $\Psi(n;p=0)$
decreases exponentially, as well as the survival probability of an
immobile particle $A$, $\Psi_0(n;p,\lambda)$, so that it may be assumed that
$\Psi(n;p,\lambda)$ also decreases exponentially with a constant larger than
$\underline{k}$  given by eq.(50), but it is difficult to estimate  this
constant precisely.

\section{Conclusion}

We have developed the stochastic lattice theory of the
annihilation kinetics of a species $A$ by another species $B$, in
systems in which the $A$ and $B$ particles perform independent,
stochastic motions which can be rather general. We obtained formal
expressions for the survival probability of $A$. This probability
cannot be evaluated exactly if $A$ is actually mobile. However, we
proved that Thu $A$ particle survival probability is always larger
in case when $A$ is immobile than when it moves. We have shown
that this the so-called Pascal principle holds for a large class
of stochastic motions, provided $B$ executes a random walk
satisfying certain reasonable assumptions. This conclusion is of a
special importance in view of its implications on chemical
reactions or population dynamics. It also allows to demonstrate
that in low dimensions the survival probability of the $A$
particle is essentially insensitive to its motion and fluctuations
of the reactivity,
 and does not obey the conventional
 mean-field laws
of
chemical kinetics.
This result is extended
to the case of stochastically-gated reactions, including the case when
the fluctuations of  reactivity are time-correlated. Furthermore, the
method used here
allows to obtain the chemical constant of a
stochastically-gated annihilation of immobile $A$ particles
 in a straightforward manner.

The stochastic analysis
of chemical reactions should be developed in different directions,
in order to consider more realistic models.
In particular, it would be interesting
to address the case
when the activity of $A$ can also fluctuate.
However, the most necessary improvement of the theory would be to extend it
to the analysis of analogous reaction kinetics in
continuous space and time.

\appendix
\section{properties of the random walk of particles $B$}

\textbf{Stochastic motion of $B$ particles.}
The constraints imposed on this stochastic process are described in
Section 2.1. We show here
that the main condition in eq.(5)
\begin{eqnarray}
\displaystyle
P(Y_n = y|Y_0 = 0)\leq P(Y_n = 0 |Y_0 = 0)\equiv R_n
\end{eqnarray}
is satisfied, if the elementary transition probability is symmetric and
obeys
\begin{eqnarray}
\displaystyle
\nonumber p(x|y) =
p(x-y) = p(y-x) \mbox{ and } p(0)\geq 1/2.
\end{eqnarray}

To show this, it is expedient to use first the well-known formula for the propagator of a random walk
on a d-dimensional regular lattice (see, e.g. Ref.\cite{hug}):
\begin{eqnarray}
\displaystyle
P(Y_n = y|Y_0 = 0) = \frac{1}{(2\pi )^d}\int_{\mathcal{B}}e^{-iz \cdot y}(\phi(z))^nd\mathbf{z},
\end{eqnarray}
where $\mathcal{B}$ is the first Brillouin zone of the lattice, while
 $\phi(z)$ is the so-called structure function
\begin{eqnarray}
\displaystyle
\phi(z) = \sum_{y}e^{iz \cdot y}p(y)
\end{eqnarray}
in which equation $z \cdot y$ stands for the scalar product of
two $d$-dimensional vectors $z$ and
$y$, while $d\mathbf{z}$ represents the differential element in a
$d$-dimensional space.

Now, according to our assumption, $p(y) = p(-y)$, which implies
that
\begin{eqnarray}
\displaystyle
\nonumber \phi(z) = \sum_{y}\cos{(z \cdot y)}p(y) = p(0) +\\
+ (1-p(0))\sum_{y\neq0}\cos{(z \cdot y)}p(y)/(1-p(0))
\end{eqnarray}
Evidently, the second term in  eq.(A.4) is bounded from above by unity
 if  $p(0)\geq 1
-p(0)$, or $p(0)\geq 1/2$.  In this case, $P(Y_n = y|Y_0 =
0)=(2\pi)^d\int_{\mathcal{B}}\cos{(z \cdot y)}(\phi(z))^nd\mathbf{z}$ is
maximal for $y = 0$, which proves the inequality in eq.(A.1).

Furthermore, one readily notices  that
if this condition is fulfilled,  $R_n$ is a decreasing function of $n$.

\textbf{Relation with first return time.}
The probability $R^1_n$ that the first return of $B$ to its
initial position occurs at time moment
$n$ is classically obtained from the relation
\begin{eqnarray}
\displaystyle
\nonumber P(Y_n = 0|Y_0 = 0) &\equiv& R_n = R^1_n + \sum_{1\leq k\leq n-1}R_{n-k}R^1_k =\\
&=& \sum_{0\leq k\leq
n-1}R_{n-k}R^1_k \mbox{ for } n\geq 1
\end{eqnarray}
where
$R_0 \equiv 1$
and $R^1_0 \equiv 0$. Then, the generating functions of  $R_n$  and  $R^1_n$ satisfy
\begin{eqnarray}
\displaystyle
\widehat{R}(s)-1=\widehat{R}(s)\widehat{R}^1(s),
\end{eqnarray}
which yields
\begin{eqnarray}
\displaystyle
\widehat{R}(s) = (1-\widehat{R}^1(s))^{-1} \to 1/S \mbox{ if } s \to 1,
\end{eqnarray}
$S$ being the probability that the $B$ particle never returns to its initial position.

\section{The case of Polya random walks}

The assumptions of section 2 exclude the Polya random walks, or any
random walk such that there is a $0$ probability to
stay immobile at each
integer time: $p(0) = 0$,
i.e. a random walk in which a particle is forced to make a move
at each integer time moment.
In this case, the probability to return to
the initial position is obviously 0 at any odd time moment, and the inequality
in eq.(5), which plays a basic role in our reasonings,
holds only at even moments of
time.
More precisely, possible displacements of the random walker in this case
can be divided into
two complementary subsets $E_0$ and $E_1$, such that the total
displacement during time $n$ necessarily belongs to $E_0$ if $n$ is
even,
and to $E_1$ if $n$ is odd.
Thus Pascal principle cannot apply in a strict sense.

We can recover the
Pascal principle for Polya random walks if we
slightly change the rules of our model,
imposing, for instance, that the $A$
particle moves only
on the lattice $E_0$,
and that the $B$ particles
are distributed on $E_0$ only.
Thus no reaction can occur at odd times,
and we only consider even times $n = 2n'$.
Then, the  inequality in eq.(5)
applies, as well as all previous calculations,
and the Pascal principle holds.

However, it is interesting to discuss the case when $B$ performs a
Polya random walk, if the evolution of particles is not restricted on a
sub-lattice.  The eq.(16),
\begin{eqnarray}
\displaystyle
\nonumber P(Y_n = x_n| Y_0 = y_0) =  \sum_{0\leq k<n} P(Y_n = x_n| Y_k = x_k)\\
P^1(k|\Gamma_A,y_0)
\end{eqnarray}
is still valid, (with possibly many vanishing terms), but the inequality
in eq.(19) cannot be deduced from it.

If, in eq.(B.1),
$n - k$ is even, $x_n - x_k$ should belong to $E_0$, and inequality in eq.(5) holds
\begin{eqnarray}
\displaystyle
P(Y_n = x_n|Y_k = x_k)\leq R_{n-k}
\end{eqnarray}

On the contrary, if $n-k$ is odd, $R_{n-k} = 0$, but we have
\begin{eqnarray}
\displaystyle
\nonumber &&P(Y_n = x_n|Y_k = x_k)=\sum_yp(x_n|y)P(Y_{n-1} = y|Y_k = x_k)\\
&\leq& \sum_yp(x_n|y)R_{n-1-k}= R_{n-1-k}
\end{eqnarray}

Consequently, from eq.(B.1) we can deduce the inequality
\begin{eqnarray}
\displaystyle
P(Y_n = x_n|Y_0 = y_0)\leq \sum_{0\leq k<n} R_{n-k}^*P^1(k|\Gamma_A,y_0)
\end{eqnarray}
where we have used the notation
\begin{eqnarray}
\displaystyle
R_{k}^*\equiv R_k+R_{k-1}^*
\end{eqnarray}

Summing both sides of eq.(B.3) over $y_0$ gives, with the same notations,
\begin{eqnarray}
\displaystyle
1\leq \sum_{0\leq k<n} R_{n-k}^*S(k|\Gamma_A)
\end{eqnarray}
Applying  the generating functions technique, we find then
\begin{eqnarray}
\displaystyle
\frac{1}{1-s}\leq \widehat{R}^*(s)\widehat{S}(s|\Gamma_A)
\end{eqnarray}
where the generating function of  $R^*_k$,  eq.(B.5),
is given by
\begin{eqnarray}
\displaystyle
\widehat{R}^*(s)=(1+s)\sum_{0\leq n\leq \infty}R_{2n}s^{2n}=(1+s)\widehat{R}(s)
\end{eqnarray}

If now the $A$ particle is fixed
at the origin, eq.(B.1) becomes
\begin{eqnarray}
\displaystyle
\nonumber P(Y_n = 0| Y_0 = y_0) &=&  \sum_{0\leq k<n} P(Y_n = 0| Y_k = 0)P^1(k|0,y_0) \\
&=&\sum_{0\leq k<n}R_{n-k} P^1(k|0,y_0)
\end{eqnarray}
Summing both sides of this equation
over $y_0$ and turning to the generating functions, we find, instead of eq.(B.7), the following equation
\begin{eqnarray}
\displaystyle
\frac{1}{1-s}= \widehat{R}(s)\widehat{S}(s|0)
\end{eqnarray}

Now, on comparing it with eq.(B.7), we infer that
\begin{eqnarray}
\displaystyle
(1-s) \widehat{S}(s|\Gamma_A) \geq \frac{1}{2}(1-s) \widehat{S}(s|0)
\end{eqnarray}
which implies
that, asymptotically,
\textit{if  $A$ moves,
the reaction integral for a given trajectory $\Gamma_A$ is not
smaller than half of
the reaction integral when $A$ is immobile}.

This unexpected conclusion requires some comments. First, it can
be noticed that the equality in eq.(B.11) can be realized in a
particular example. In fact, assume that $B$ performs a classical
Polya  random walk on a $d$-dimensional lattice: at each step it
can jump with an equal probability $1/(2d)$ at one of the
neighboring sites. Now, we choose a special trajectory $\Gamma_A$
for $A$, consisting of
 jumps from the origin 0 to one of its nearest neighbors, 1, and
returns:  thus, at each even time $n$, $A$ is at 0, whereas $A$ is at 1 at each odd time.

For such a trajectory eq.(B.1) reads
\begin{eqnarray}
\displaystyle
P(Y_n = x_n| Y_0 = y_0) =  \sum_{0\leq k<n} R^{\circ}_{n-k}P^1(k|\Gamma_A,y_0)
\end{eqnarray}
where
\begin{eqnarray}
\displaystyle
\nonumber R^{\circ}_{n-k}=R_{n-k}=P(Y_n = 0| Y_k = 0)=P(Y_n = 1| Y_k = 1) \\
\nonumber \mbox{ if $n-k$ is even} \\
\nonumber = R_{n-k+1}=P(Y_n = 1| Y_k = 0)=P(Y_n = 0| Y_k = 1) \\
\nonumber \mbox{ if $n-k$ is odd}
\end{eqnarray}
since
\begin{eqnarray}
\displaystyle
\nonumber &&R_{2n}=P(Y_{2n} = 0| Y_0 = 0)=(2d)^{-1} \times \\
&\times&\sum^{'} P(Y_{2n-1} = y| Y_0 = 0)=P(Y_{2n-1} = 1| Y_0 = 0),
\end{eqnarray}
where the prime designates that we sum only over the nearest
to the origin sites.
Consequently, we can write
\begin{eqnarray}
\displaystyle
R^{\circ}_{n-k}=R_{n-k}+R_{n-k+1}
\end{eqnarray}

Summing eq.(B.13) over $y_0$
we obtain for the generating functions
\begin{eqnarray}
\displaystyle
\frac{1}{1-s}=\widehat{R}^{\circ}(s)\widehat{S}(s|\Gamma_A)
\end{eqnarray}
with
\begin{eqnarray}
\displaystyle
\widehat{R}^{\circ}(s)=\frac{1+s}{s}\widehat{R}(s)
\end{eqnarray}
Thus, for this special trajectory, we actually obtain
\begin{eqnarray}
\displaystyle
(1-s) \widehat{S}(s|\Gamma_A) = \frac{1}{2}(1-s) \widehat{S}(s|0) \mbox{ if
} s \to 1
\end{eqnarray}
which implies that the reaction is twice slower
than for an immobile $A$. For instance, in three
 dimensions, the expression in eq.(B.16) is
bounded when $s \to 1$,
which means that
if $A$ moves according to the previous rules,
the reaction constant in case of a mobile $A$
is half of
the reaction constant for an immobile $A$.
If an average is taken over trajectories, the
Pascal principle can be  valid or not,
depending on the probability weight of the different trajectories.

The physical reason of this conclusion may be understood if we
consider the relative motion with respect to $A$. It is seen that
at each integer time the relative displacement of $B$ in the
direction 0-1 can be 0 or 2 at odd times, and 0 or -2 at even
times. On the other hand, the reaction integral at time $n$ is
related to the number of distinct sites visited by $B$ up to time
$n$. This number is clearly lower if $A$ moves according to the
foregoing rules, than if $A$ is immobile, which explains that the
reaction is slower in the first case. The same behavior can occur
each time $A$ and $B$ are performing Polya walks with the same
jump times.

However, it should be pointed out that when $A$ and $B$ both
perform Polya random walks with the same jump times, it may well
happen that they exchange their positions during simultaneous
jumps. In this case, they do not react according to the rules we
used here, but it can be relevant to adopt different rules,
 depending on the actual phenomenon to be modelled. Then, the results could depend
very much on these rules.

\end{document}